\documentstyle[aps,manuscript]{revtex}
\begin{document}
\draft
\title{Discreteness-Induced Oscillatory Instabilities of
Dark Solitons}

\author{Magnus Johansson$^1$ and Yuri S. Kivshar$^2$}

\address{$^1$ Department of Physics and Measurement Technology,
Link{\"o}ping University, S-581 83 Link{\"o}ping, Sweden \\
$^2$Optical Sciences Center, Research School of Physical Sciences and
Engineering, The Australian National University, Canberra ACT 0200,
Australia}

\maketitle

\begin{abstract}
We reveal that even weak inherent discreteness of a nonlinear model can lead
to instabilities of the localized modes it supports. We present the first
example of an oscillatory instability of dark solitons, and analyse how it
may occur for dark solitons of the discrete nonlinear Schr\"odinger and
generalized  Ablowitz-Ladik equations.
\end{abstract}


\newpage

Wave instabilities are probably the most remarkable nonlinear
phenomena that may occur in nature \cite{infeld}. One of the first
instabilities discovered for nonlinear models was the modulational instability,
which is known to be an effective physical mechanism in fluids \cite{mi_fluid}
 and optics \cite{has} for break-up of continuous modes into solitary waves.
Also, the solitary waves themselves may become unstable, and the analysis
of their instabilities is an important problem of nonlinear physics.
Instabilities
are known to occur for both bright \cite{bright_review} and dark
\cite{dark_review} solitary waves of different {\em nonintegrable}
nonlinear models.

Recently, a new type of solitary-wave instability, {\em oscillatory
instability}, has been found to occur for bright Bragg gap solitons in the
generalized Thirring model \cite{osc_bar}. Such an instability is
characterized by complex eigenvalues, and its scenario is associated with a
resonance between the long-wavelength radiation and soliton internal modes
which appear in the soliton spectrum when the model becomes nonintegrable
\cite{internal}. In spite of the fact that oscillatory instabilities appear
often in dissipative models \cite{dissip}, their manifestation in
continuous Hamiltonian models is
rare \cite{weinstein}, and so far {\em no example has been known for
oscillatory instability of dark solitons}.

The aim of this Letter is twofold. First, we analyse what we believe
to be the first examples of oscillatory instabilities of dark solitons, by
considering the important cases of the discrete nonlinear Schr\"odinger
(DNLS) and
generalized Ablowitz-Ladik (AL-DNLS) models.  We reveal  {\em two different
scenarios} for the dark-soliton oscillatory instability, which may occur
due to either a resonance between radiation modes and the soliton internal
mode, or a resonance between two soliton internal modes. Second, we
demonstrate that even a weak inherent discreteness may drastically modify
the dynamics of  a nonlinear system leading to instabilities which have no
analog in the continuum limit.

First, we consider the well-known DNLS equation,
\begin{equation}
\label{DNLS}
i \dot{\psi_n} + C(\psi_{n+1} + \psi_{n-1}) + | \psi_n|^2 \psi_n = 0,
\end{equation}
where the dot stands for the derivative in time. Stationary localized
solutions of Eq. (\ref{DNLS}) in the form
$\psi_n (t)= \phi_n e^{i \Lambda t}$,  where $\Lambda = 2 C \cos k +
(\phi^{(0)})^2$, may exist as dark-soliton modes with the
nonvanishing boundary conditions $\phi_n \rightarrow \pm \phi^{(0)} e^{i k
n }$  ($n  \rightarrow \pm \infty $), provided the background wave is
modulationally stable, i.e. for $C \cos k < 0$ \cite{peyrard}.  Without
loss of generality, we can put the background intensity to unity,
$\phi^{(0)}=1$.

The structure of the dark-soliton modes of Eq. (\ref{DNLS}) has been discussed
earlier \cite{kkc,hennig}. Here, we consider the case $C>0$, in which the
background wave has $k=\pi$ and is `staggered', as shown in Figs. 1(a,b).
The transformation $\psi_n \rightarrow (-1)^n \psi_n$ immediately yields
the corresponding `unstaggered' modes ($k=0$) for negative $C$.

The dark-soliton modes presented in Figs. 1(a,b) describe two types of
stationary `black' solitons \cite{dark_review} in a discrete lattice, the
on-site mode (A-mode) centered with zero intensity at a lattice site, and
the inter-site mode (B-mode)  centered between two sites. These two modes
can be uniquely followed from the  continuous limit ($C \rightarrow
\infty$) to the `anticontinuous' limit ($C=0$).  At  $C=0$, the A-mode
takes the form $\phi_n =  (...-1,+1,-1,0,+1,-1,+1,...)$, and it describes a
single  `hole' in a background wave with constant amplitude and a $\pi$
phase shift across the hole.  Similarly, the B-mode takes the form  $\phi_n
= (...-1,+1,-1,-1,+1-1,...)$, and it describes the lattice oscillation mode
with  a $\pi$ phase shift between two neighboring sites and no hole.

Linear stability of the A-mode for small enough $C$ follows from Aubry's
theorem (Theorem 9 in Ref. \cite{A}) relating the linear stability of
multi-breathers  to the extrema of the {\it effective action}
which is a function  of the relative phases $\alpha_n$ of the single
breathers.  For the DNLS equation (\ref{DNLS}) with a positive
nonlinearity, {\em local minima} of the  effective action correspond to
stable solutions. A perturbative expression of  the effective action to
order $C^2$ was obtained in Eq. (A10) in \cite{JA}. For  small $C$, it is
enough to consider the phase-interactions between nearest and
next-nearest neighboring sites. Then, the lowest order  contribution to the
effective action from neighboring excited sites is $\sum_n [2 C| \Lambda|
\cos(\alpha_{n+1}-\alpha_n) - C^2 \cos ^2  (\alpha_{n+1}-\alpha_n)]$, while
the contribution from the next-nearest-neighbor  interaction is $-C^2
\sum_{n \neq n_0} \cos ( \alpha_{n+1}-\alpha_{n-1}) +  2 C ^ 2 \cos (
\alpha_{n_0+1}-\alpha_{n_0-1})$, where $n_0$ is the site with the  `hole'.
For the A-mode, $\alpha_{n+1}-\alpha_n = \pi$, $\alpha_{n+1}-\alpha_{n-1}
= 0$, $n \neq n_0$, $\alpha_{n_0+1}-\alpha_{n_0-1}=\pi$, so that the
effective  action has a local minimum and the mode is stable for
small $C$.

For the B-mode, there are two neighboring sites having the same phase at
the  center, which suggests that the corresponding extremum of the
effective  action is a saddle. Numerical investigation of the eigenvalue
problem also shows that the B-mode is unstable
for all $C$.

Let us consider the linear stability of the dark-soliton modes for
nonvanishing $C$. Assuming $\psi_n (t) = [\phi_n + \epsilon_n (t)] e^{i
\Lambda t }$, we obtain the  linearized equations for $\epsilon_n$
\[
i
\dot\epsilon_n+C(\epsilon_{n+1}+\epsilon_{n-1})+2|\phi_n|^2\epsilon_n+\phi_n^2
\epsilon_n^\ast - \Lambda \epsilon_n = 0.
\]
Writing $\epsilon_n=\xi_n+i\eta_n$ for real $\phi_n$ yields
\[
\frac{d}{dt}\left(\begin{array}{c} \xi_n \\ \eta_n \end{array} \right) =
\left(\begin{array}{cc} 0 & H^+ \\ -H^- & 0 \end{array}
\right)\left(\begin{array}{c} \xi_n \\ \eta_n \end{array} \right) \equiv
\hat{M}
\left(\begin{array}{c} \xi_n \\ \eta_n \end{array} \right),
\]
where for a system of $N$ sites $\hat{M}$ is a $2N \times 2N $ matrix and
$H^+$ and  $H^-$ are $N \times N $ matrices with time-independent
coefficients, $H_{ij}^\pm  = [ \Lambda - ( 2 \mp 1 ) \phi_i^2 ]
\delta_{i,j} - C  (\delta_{i,j+1}+\delta_{i,j-1}) $ (boundary conditions
not explicitly taken into  account). Linear stability is then equivalent to
the matrix $\hat{M}$ having all its  eigenvalues on the imaginary axis.

When $C=0$, all eigenvalues of $\hat{M}$ lie at zero except, for the
A-mode, one complex conjugated pair at $\pm i $ corresponding to a mode
localized on the  `hole'.  When $C$ is increased, the eigenvalues at zero
spread on the  imaginary axis creating {\em a phonon band of extended
states}, which corresponds to a  continuous spectrum for large $N$.
Assuming $\epsilon_n  = a e^{i(\kappa n - \omega t) } + b e^{-i(\kappa n -
\omega t) } $ yields the  dispersion relation,
$\omega = \pm \sqrt { 16 C^2 \cos^4 (\kappa/2) + 8 C  \cos^2 (\kappa/2)}$,
so that the eigenvalues corresponding to the continuous
band lie between 0 (at $\kappa=\pi$) and $\pm i \sqrt{16 C^2+8C}$ (at
$\kappa=0$). On the  other hand, the pair of eigenvalues at $\pm i $ will
move towards zero on the  imaginary axis as $C$ is increased. For small
$C$, we can assume that the
corresponding eigenmode is almost completely localized at the hole, so that
$\epsilon_{n_0+1}\approx \epsilon_{n_0-1} \approx 0$. Since $\phi_{n_0}=0$,
we  obtain that this mode oscillates with frequency $\Lambda = 1-2C$, so
that the  corresponding eigenvalues are $\pm i(1-2C)$, to the lowest order
in $C$. Then,  equating this result with the expression above for the edge
of the continuous band gives an estimate to the value of $C$ where these
two eigenvalues coincide: $C=1/\sqrt{3} - 1/2 \approx 0.07735 $. This
agrees  quite well with the exact, numerically obtained value $C \equiv
C_{\rm cr} \approx  0.07647$. At $C=C_{\rm cr}$, {\em a Hopf-type bifurcation}
occurs, as two complex conjugated pairs of eigenvalues leave the imaginary
axis and go out in  the complex plane (see Fig. 2). Thus, {\em an oscillatory
instability} occurs for the  A-mode when $C>C_{\rm cr}$, with the
instability growth rate $\lambda$ given by the real part of  the unstable
eigenvalue (see Fig. 2). For the B-mode, all eigenvalues lie at zero when
$C=0$. As  soon as $C$ is increased, one pair go out on the real axis and
stay there  for all $C>0$. {\em Thus, the B-mode is always unstable.}

For $C > C_{\rm cr}$, the imaginary part of the unstable
eigenvalue moves  towards zero (i.e, the oscillation frequency
decreases) as $C$ is increased. The real part of the eigenvalue vs. $C$
is shown in Fig. 2 for two different system sizes, and compared with
estimations of the maximum growth rate from direct numerical integration of
the DNLS model for chains large enough to eliminate the influence of the
boundaries. In all cases the  instability is largest for $C \approx 0.32$
where $\lambda \approx 0.120 $. For  larger $C$, there appear stable `windows'
for the finite systems, where the mode is stabilized by the boundaries. The
location of these windows depend  critically on the size of the system and
the boundary conditions (in Fig. 2, periodic boundary conditions
$\psi_{N+1}=\psi_1$ are used).

Similar `re-entrant instabilities' have been found for discrete breathers
of the finite-width Klein-Gordon chains \cite{MA}. Qualitatively, the
explanation is  that for small systems, the eigenvalues
corresponding to the `continuous' spectrum are rather sparsely distributed,
and the corresponding eigenvectors are localized over the size of the
system. Thus, as the eigenvalue corresponding to the unstable mode
approaches the imaginary axis when increasing $C$, it may find a `hole' in
the spectrum and join the imaginary
axis for a while. Then, a further increase of $C$ causes a collision with
the next imaginary eigenvalue, and a new instability occurs. This
procedure repeats itself until the eigenvalue has passed the phonon band
eigenvalue which is closest to zero. After this the state is stable  for
all larger values of $C$ (e.g., in Fig. 2 the state is stable for all
$C>1.11$ when $N=61$, and for all $C>1.38$ when $N=121$). Increasing the
system  size implies that the eigenvalue distribution on the imaginary axis
will be more dense, so that the stable windows will be smaller  and {\em
completely disappear in the limit of the infinite system}, where the
unstable eigenvalue never joins the imaginary axis.

For an infinite DNLS chain, the instability growth rate decreases in an
exponential-like way to zero for large $C$, and thus indicates  that the
dark mode is unstable for all $C>C_{\rm cr}$.  A quite  good asymptotic fit
is obtained with a stretched exponential, $\lambda \sim \exp (-bC^\gamma)$,
with $\gamma \approx 0.7 $ (see inset in Fig. 2).  We also investigated the
case of varying
$\phi^{(0)}$ to keep the {\it complementary norm} \cite{dark_review}
constant when varying $C$.
In that case, we found the  asymptotic decay of the instability growth rate
with $C$ to be
faster than purely  exponential.

Figure 3 shows the time evolution for two different values of $C$
when the initial state is a slightly perturbed dark A-mode. A small
perturbation has been chosen to be approximately in the direction of the
unstable eigenvector. If instead a random perturbation is chosen, the
qualitative behaviour is the same, except that there will be an initial
small amount of radiation before the unstable internal eigenmode gets
excited. The unstable eigenvector is always  spatially symmetric around the
central site, and therefore antisymmetric w.r.t  the dark mode itself,
since the latter is antisymmetric.  Thus, the main effect of the
instability is a transition from a black (zero intensity at the middle) to
grey (nonzero intensity at the middle) soliton, as can be clearly
identified at least for $C>0.3$. This is similar to the instability
scenario of dark solitons in some continuous models \cite{dark_review},
except the oscillatory dynamics.  The direction of the resulting grey
soliton is determined by the sign of the perturbation projected on the
unstable eigenmode. For large $C$,  the grey soliton is almost black, moves
slowly, and the radiation is small, while for smaller $C$ the
minimum value of the soliton intensity increases as well as its velocity
and the amount of radiation. For small and intermediate values of the
coupling parameter $C$, the resulting grey soliton decays continuously into
radiation (see Fig. 3).

It is important to study the oscillatory instability of dark solitons for
other types of nonlinear lattices. Here, we consider the AL-DNLS equation
$i \dot{\psi_n} + C ( \psi_{n+1} + \psi_{n-1} ) + \mu | \psi_n|^2 \psi_n +
\frac{1}{2}(\mu-1) |\psi_n|^2  ( \psi_{n+1} + \psi_{n-1} ) = 0$,  where $0
\leq \mu \leq 1$. The case $\mu = 0$ corresponds to the integrable AL
model, whereas the case $\mu = 1$, to the DNLS model analyzed above.

With the general form  $\psi_n (t)= \phi_n e^{i \Lambda t }, $
and boundary conditions  $\phi_n \rightarrow \pm \phi^{(0)}
e^{i k n}$ ($n  \rightarrow \pm \infty$), the frequency $\Lambda$ is
determined by $\Lambda = 2 C \cos k + (\phi^{(0)})^2 [\mu + (\mu-1) \cos
k]$. In particular, for the black mode ($k = \pi $) the relation becomes
$\Lambda = - 2 C + (\phi^{(0)})^2$,  just as for the DNLS model. As above,
we put $\phi^{(0)}=1$ without loss of generality.

In the AL-DNLS model, there is a lower limit of $C$ for the existence of
dark solitons, due to the instability of the background when the effective
coupling changes sign (see, e.g., Ref. \cite{KS}). This  occurs when $C+
(\mu-1)(\phi^{(0)})^2/2 = 0 $, so that, when $\phi^{(0)}=1$, the dark modes
exist only for $C > (1-\mu)/2$ (Fig. 4, dashed line).

Considering $\psi_n (t) = [\phi_n + \epsilon_n (t)] e^{i \Lambda t}$ as above,
we obtain the linearized equations for the small perturbation $\epsilon_n$,
and the dispersion relation for the continuous spectrum
(for the `staggered' mode with $\lim_{|n| \rightarrow \infty} \phi_n^2 = 1
$): $\omega = \pm 2 \sqrt { \nu(\kappa) [1 + \nu (\kappa)]},$
where $\nu(\kappa) \equiv (2 C + \mu - 1 ) \cos^2(\kappa/2)$.
Therefore, the eigenvalues corresponding to the continuous band lie between
0 (at $\kappa=\pi$) and $\pm i 2 \sqrt{(2 C + \mu -1) (2C+\mu)}$ (at
$\kappa=0$) for  $C > (1-\mu)/2$.

As above, an approximate expression for the onset of instability can be
obtained by assuming that the internal mode is completely localized at the
hole, so that $\epsilon_{n_0+1}\approx \epsilon_{n_0-1} \approx 0$. Then,
the corresponding eigenvalues are $\pm i(1-2C) $, to the lowest order in
$C$. The collision with the phonon band edge  is then expected to occur for
$C \approx (1-4\mu)/6 + \sqrt{\mu^2+\mu+1}/3$.  Comparing  this result with
the numerically obtained values of $C$ for the bifurcation (see Fig. 4)
shows that the instability for smaller $\mu$ occurs for smaller $C$ than
expected from this analytical estimate. The reason why the bifurcation
occurs earlier is, that it in fact results not from a collision with the
band edge phonon, but with a second localized mode that has bifurcated from
the band edge slightly before (see the insets in Fig. 4). As a matter of
fact, studying the bifurcation for the DNLS model very closely shows that
also in this case the collision probably occurs with a second localized
mode coming from the phonon band. However, this localized mode only occurs
extremely close to the bifurcation when $\mu$ is close to 1 (for $\mu=1$ we
observe it for $C>0.076464$, and the bifurcation occurs at
$C=0.076468...$). And since it will be very weakly localized, we are not
able to say for sure whether it will be localized or not for the infinite
system. But for smaller $\mu$, this localized mode undoubtedly exists, and
has the same spatial symmetry as the `hole' mode. It describes an internal
degree of freedom of the dark soliton, i.e.  its {\em internal mode}.

Otherwise, the qualitative scenario with stable and unstable windows for
finite systems and instability all the way up to the continuum limit for
the infinite system remains the same for all AL-DNLS lattices with $0<\mu
\leq 1$. The exception is the limit of the AL model ($\mu=0$) which is
known to be {\em exactly integrable}. At $\mu =0$, dark solitons are always
stable for all $C$ and their spectrum has a pair of eigenvalues exactly at
zero corresponding to the exact translational invariance.

In conclusion, we have described, for the first time to our knowledge, the
oscillatory instability of dark solitons. This new type of dark-soliton
instability appears due to inherent discreteness of a nonlinear lattice
model, and the universality of the instability scenario suggests that it
should be also observed in other nonlinear models supporting dark
solitons. We remark that the instability observed here may be regarded as an 
extension of the instabilities existing for small lattices \cite{eilbeck}. 

M. J. acknowledges support from the Swedish Natural Science Research Council.
Yu. K. acknowledges a partial support from the Australian Research
Council.  The results have been reported at the CECAM Workshop on Nonlinear
Localized Excitations in Condensed Matter and Molecular Physics (Lyon, July
7-10, 1998).

\begin{figure}
Fig. 1. (a),(b) Two types of `staggered' dark-soliton modes in a lattice.
Shown are the oscillation amplitude at each site (upper row) and the value
$|\psi_n|^2$ (lower row).
\end{figure}

\begin{figure}
Fig. 2. Real part of the unstable eigenvalues as a function of $C$
in the model (\ref{DNLS}) for $N=61$ (dotted) and $N=121$ (dashed) sites,
and the growth rate (solid) for an infinite system, calculated for lattices
of up to 15000 sites. Inset: asymptotic fit of the growth rate (points)
with stretched exponential (line) (see text). Below: instability scenario
shown as the evolution of
the eigenvalues of $\hat{M}$  for $N=121$ sites.
\end{figure}

\begin{figure}
Fig. 3. Time-evolution of $|\psi_n|^2$ with slightly perturbed dark
A-modes as initial conditions for $C=0.30$ (left) and $C=0.75$ (right). Lower
figures show the detailed dynamics for a few sites around the center.
\end{figure}

\begin{figure}
Fig. 4. Solid: Numerically obtained instability threshold $C_{\rm cr}(\mu)$
for the AL-DNLS model.  Dashed: Minimum value of $C$ for the existence  of
the dark localized modes.  Insets: Examples of the eigenvalues at
$\mu=0.2$, below (A) and above (B)
the threshold curve $C_{\rm cr}(\mu)$.
\end{figure}

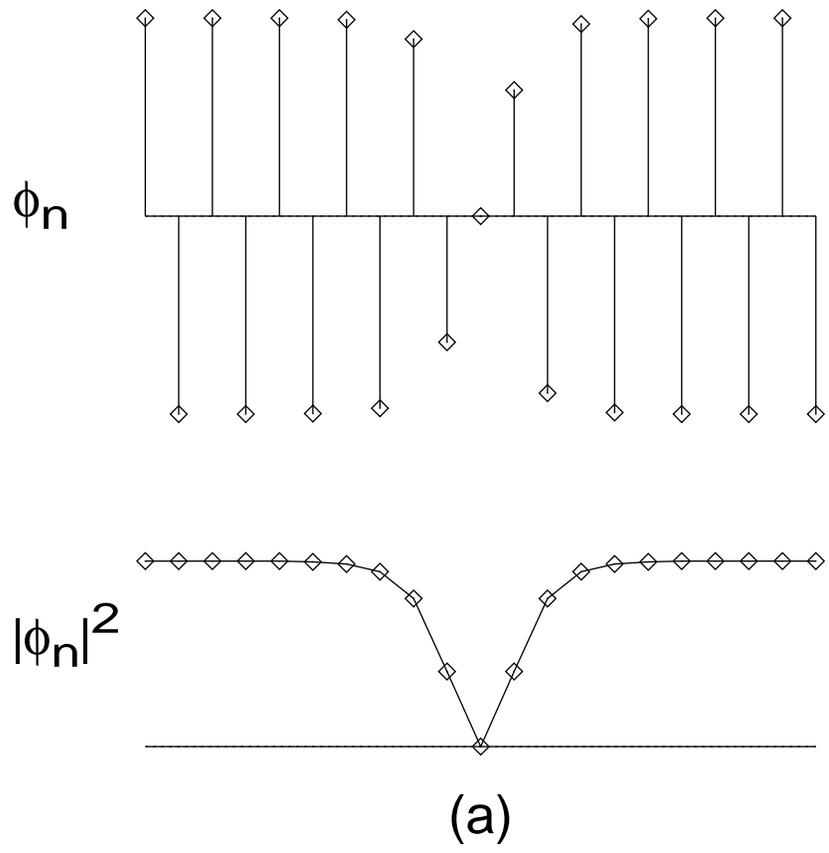
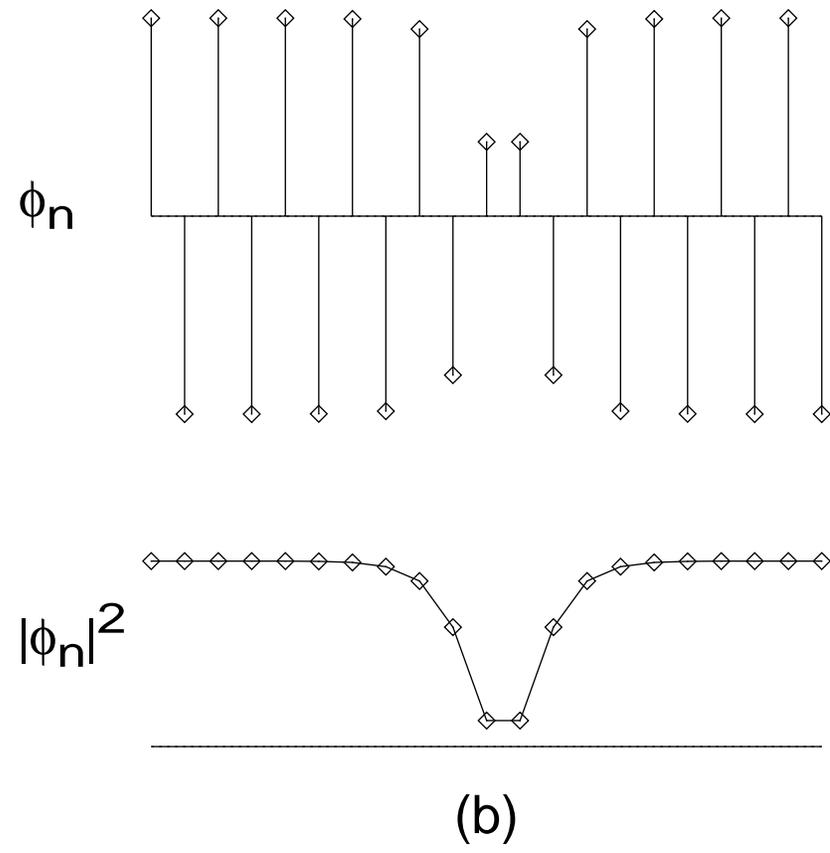
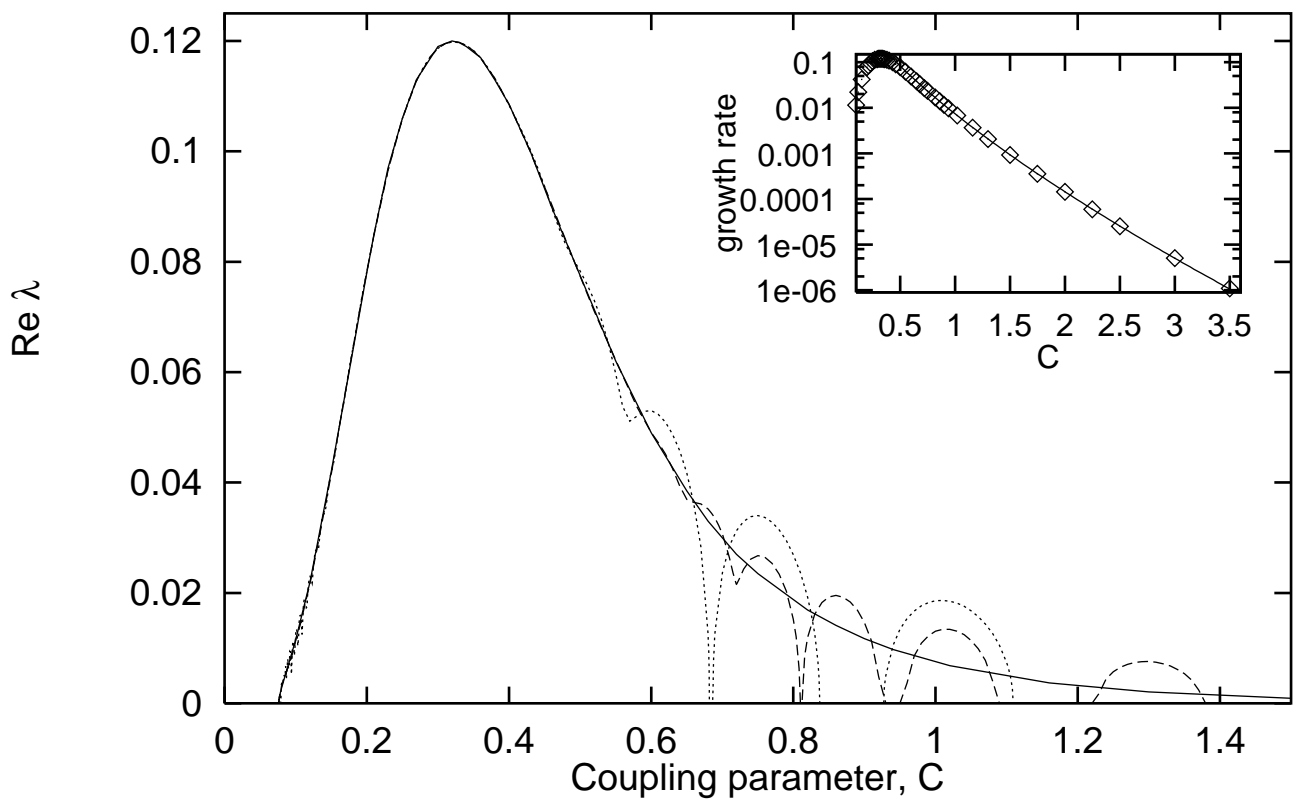
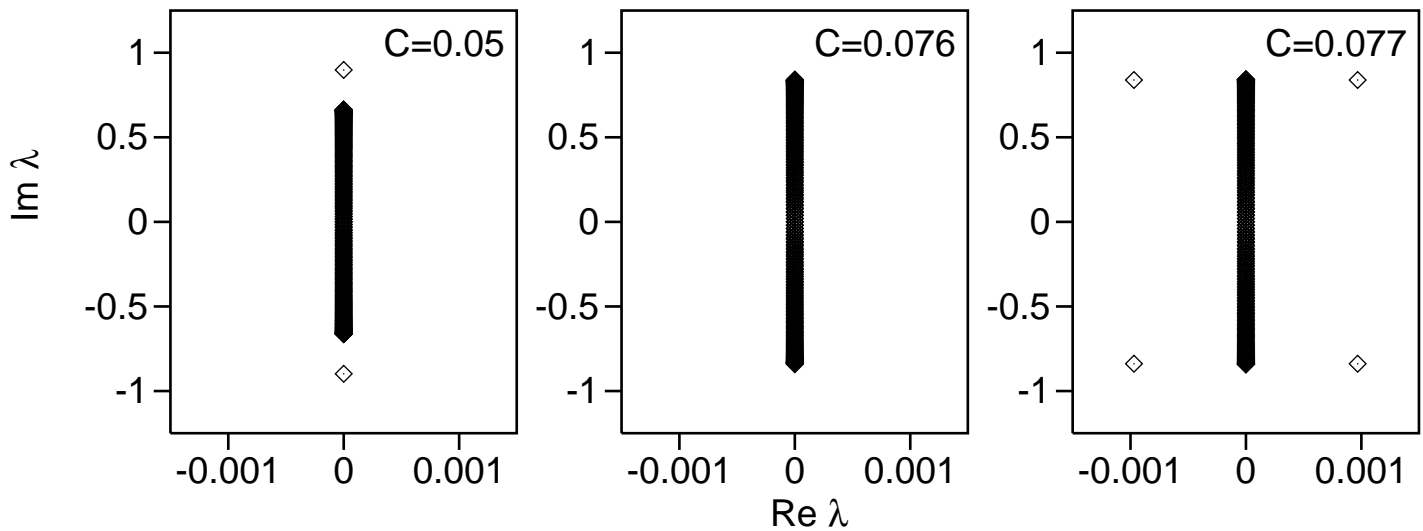
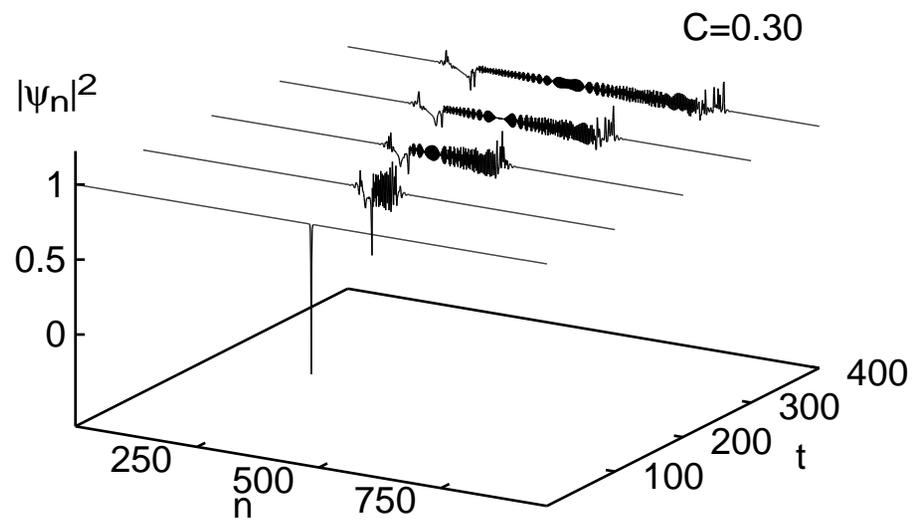
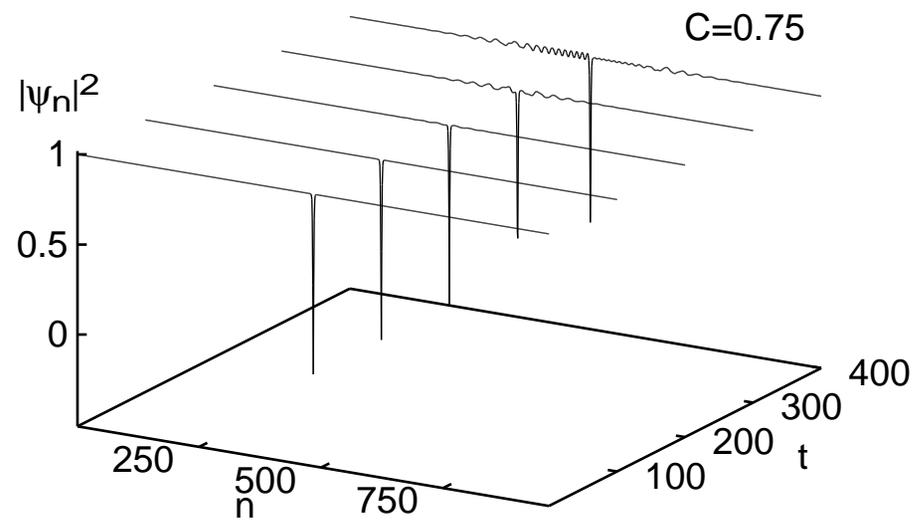
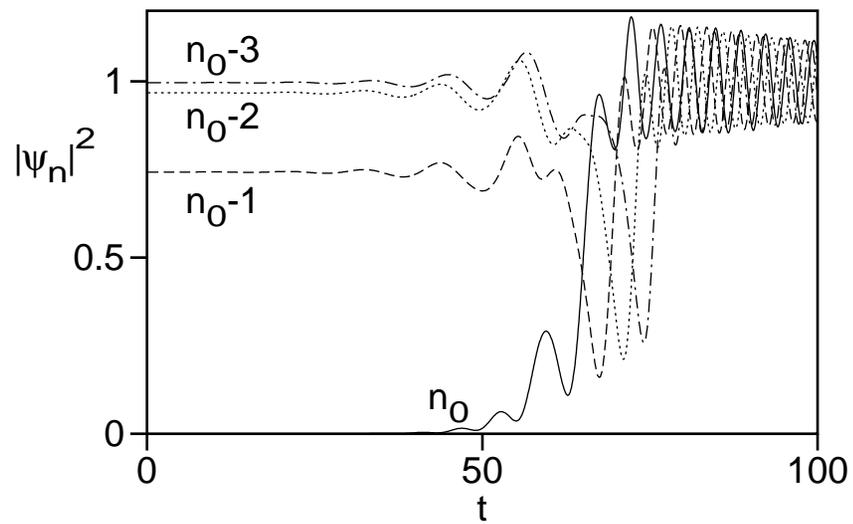
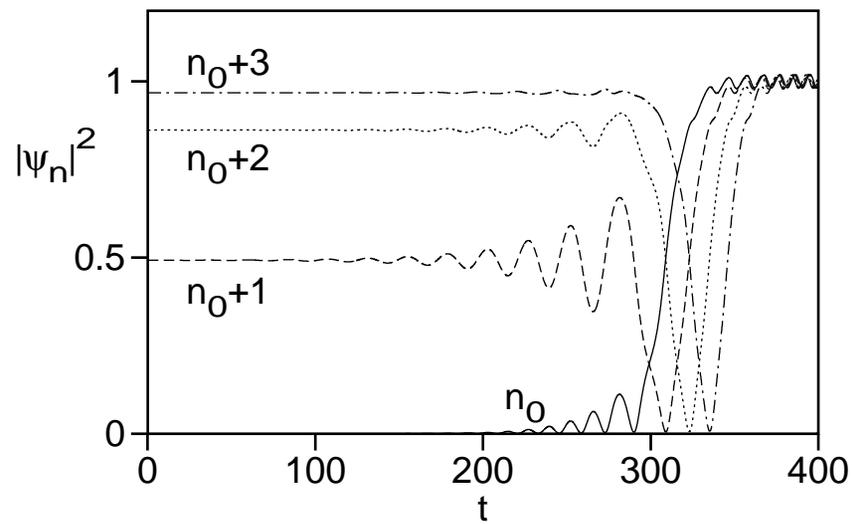
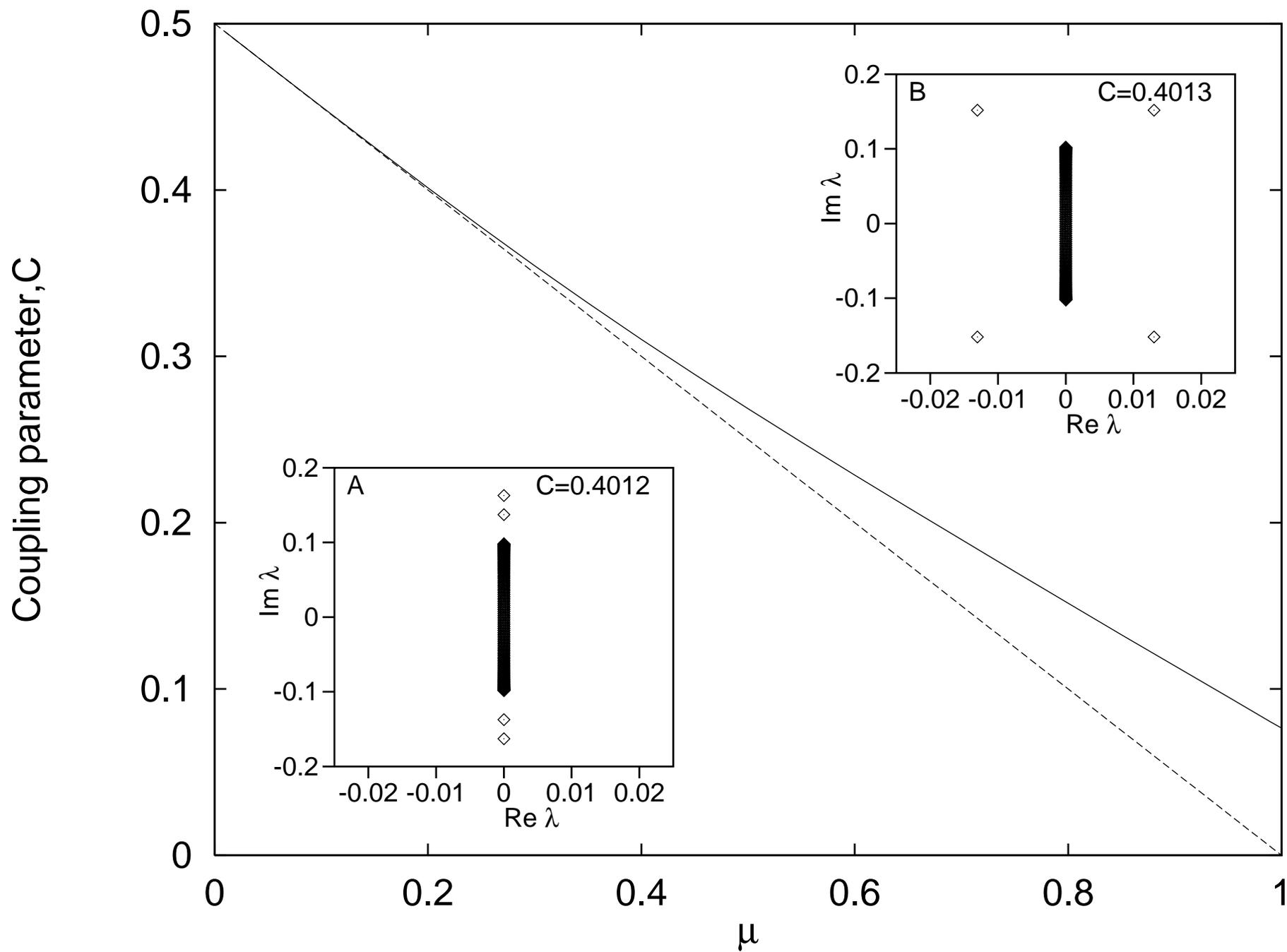
\end{document}